\documentclass[10pt]{article}
\usepackage[superscript,sort]{cite}

\usepackage{ifthen,ifpdf}
\ifpdf
	\usepackage[pdftex]{hyperref}	\hypersetup{colorlinks=true,linkcolor=black,citecolor=black,urlcolor=blue,backref=page,bookmarks=true,breaklinks=true,plainpages=false }
	\usepackage[pdftex]{graphicx}
	\usepackage[usenames,pdftex]{color}
\else
	\usepackage[colorlinks=true,breaklinks=true,plainpages=false]{hyperref}
	\usepackage{graphicx}
	\usepackage[usenames]{color}
\fi

\usepackage{amsthm,amscd,amsxtra,amsfonts,amsmath,amssymb,multirow}
\usepackage{wrapfig}
\usepackage[footnotesize]{caption}
\usepackage[tiny,compact]{titlesec}
\usepackage[textwidth=0.8in,textsize=footnotesize]{todonotes}
\usepackage{algorithm,algorithmic,extarrows}
\usepackage{bm}

\begin{document}

\title{Weighted persistent homology for osmolyte molecular aggregation and hydrogen-bonding network analysis}

\author{
D Vijay Anand$^1$,
Kelin Xia$^{1,2}$\footnote{ Address correspondences  to Kelin Xia. E-mail:xiakelin@ntu.edu.sg},
Yuguang Mu$^{1,2}$
\\
%\address{
$^1$Division of Mathematical Sciences, School of Physical and Mathematical Sciences, \\
Nanyang Technological University, Singapore 637371\\
$^2$School of Biological Sciences, Nanyang Technological University, Singapore 637371
}

\date{\today}
\maketitle

\begin{abstract}
It has long been observed that trimethylamin N-oxide (TMAO) and urea demonstrate dramatically different properties in a protein folding process. Even with the enormous theoretical and experimental research work of the two osmolytes, various aspects of their underlying mechanisms still remain largely elusive. In this paper, we propose to use the weighted persistent homology to systematically study the osmolytes molecular aggregation and their hydrogen-bonding network from a local topological perspective. We consider two weighted models, i.e., localized persistent homology (LPH) and interactive persistent homology (IPH). Similar to our previous works, we use persistent Betti number (PBN) and persistent entropy (PE) to quantitatively characterize the topological features from LPH and IPH. More specifically, from the localized persistent homology models, we have found that TMAO and urea have very different local topology. TMAO shows local network structures. With the concentration increase, the circle elements in these networks show a clear increase in their total numbers and a decrease in their relative sizes. In contrast, urea shows two types of local topological patterns, i.e., local clusters around 6 \AA~ and a few global circle elements at around 12 \AA. From the interactive persistent homology models, it has been found that our persistent radial distribution function (PRDF) from the global-scale IPH has same physical properties as the traditional radial distribution function (RDF). Moreover, PRDFs from the local-scale IPH can also be generated and used to characterize the local interaction information. Other than the clear difference of the first peak value of PRDFs at filtration size 4\AA, TMAO and urea also shows very different behaviors at the second peak region from filtration size 5\AA~ to 10 \AA. These differences are also reflected in the PBNs and PEs of the local-scale IPH. These localized topological information has never been revealed before. Other than the osmolyte systems, our weighted persistent homology models can also be used to analyze any types of networks, graphs, molecular structures, etc.
\end{abstract}

\newpage

%{\setcounter{tocdepth}{5} \tableofcontents}

%\newpage

\section{Introduction}
Tri-methylamine N-oxide (TMAO) and urea are organic osmolytes widely existing in the animal metabolisms. Deep-sea organisms use the protein stabilizing effects of TMAO to counteract the high pressure perturbation, while mammalian kidneys use the strong denaturant function of urea\cite{sahle2016influence}. As a protecting osmolyte, TMAO can counteract the urea protein-denaturing effects. Currently, it is well accepted that urea acts by directly binding to the protein backbones and side chains\cite{ganguly2017trimethylamine}. It has very litter disturbance to the surrounding water structures. TMAO's stabilization is not well understood. It has been suggested that TMAO form complexes with two to three water molecules, and protein stabilization is the result of depletion effects associated with unfavorable interaction of TMAO with protein backbone\cite{hunger2015water}. Others suggest that TMAO interacts with polypeptides and protein stablization is a results of sufactact-like effects of TMAO\cite{liao2017trimethylamine}. The interaction between TMAO and urea is also not well understood\cite{ganguly2016hydrophobic}. Even though it is suggested that the interaction is through the TMAO's modification of urea-water structures, recent experiments show that the addition of TMAO induces blue shifts in urea's H-N-H symmetric bending modes, indicating the direct interactions between the two cosolvents\cite{zetterholm2018noncovalent,xie2018large}. Although great progress has been made in both experimental and theoretical research for urea and TMAO\cite{rossky2008protein,idrissi2010effect,bandyopadhyay2014molecular,rezus2006effect,panuszko2009effects,baskakov1999trimethylamine,baskakov1998forcing,uversky2001trimethylamine,tseng1998natural,rosgen2012volume,paul2007structure,ganguly2015mutual,meersman2011x}, the detailed mechanism for their molecular aggregations and corresponding hydrogen-bonding networks still remain elusive.

Theoretically, graph or network based models\cite{radhakrishnan:1991graph, dos:2004topology,Oleinikova:2005formation}, especially the spectral graph models and combinatorial graph models, play the key role in the characterization of biomolecular structures, interaction networks, hydrogen-bonding network, etc\cite{Bako:2008water,da:2011hydrogen,Bako:2013hydrogen,choi2018graph,Choi:2014ionII,choi:2015ion,choi2016ion}. The most commonly-used graph-based measurements include, node degree, shortest path, clique, cluster coefficient, closeness, centrality, betweenness, Cheeger constant, modularity, graph Laplacian, graph spectral, Erd\H{o}s number, percolation information, etc. Differential geometry tools\cite{petvrek2007mole,edelsbrunner2005geometry,chalikian1998thermodynamic,cazals2006revisiting,Bates:2008,XFeng:2013b,KLXia:2014a,smolin2017tmao}, such as Voronoi diagram, alpha shape, geometric flows, etc, have also been considered to quantitatively characterize biomolecular structure, surface, volume, cavity, void, tunnels, interface, etc. Recently, a new topological model known as persistent homology has demonstrated a great promise in biomolecular structure, flexibility, dynamics and function analysis\cite{KLXia:2014c, KLXia:2015a,BaoWang:2016a,KLXia:2015c,KLXia:2015b}. Persistent homology based machine learning and deep learning models\cite{pun2018persistent} have achieved great successes in protein-ligand binding affinity prediction\cite{cang:2017topologynet,cang:2017integration,nguyen:2017rigidity}, protein stability change upon mutation\cite{cang:2017analysis,cang:2018representability} and toxicity prediction\cite{wu:2018quantitative}. These topology based machine learning models have constantly achieved some of the best results in D3R Grand challenge \cite{nguyen2018mathematical}. Motivated by the great success of topological modeling in biomolecules, we have applied persistent homology in the analysis of ion aggregations and hydrogen-bonding networks \cite{xia2018persistent}. The two types of ion aggregation models, i.e., local clusters and extended ion networks, can be well characterized by our model. Further, we have identified, for the first time, two types topology for the two hydrogen-bonding network systems \cite{xia2018persistent}. More recently, we study the osmolyte molecular aggregation and their hydrogen-bonding networks. Two osmolytes, i.e., TMAO and urea, are found to share very similar topological patterns with the two types of ion systems, i.e., KSCN and NaCl. Particularly, the topological fingerprints for the hydrogen-bonding network from ion systems and osmolyte systems share a great similarity. This indicates that our topological representation can characterize certain intrinsic difference between ``structure making" and ``structure breaking" systems.

More recently, weighted persistent homology (WPH) models have been proposed to incorporate physical, chemical and biological properties into topological modeling\cite{meng2019weighted}. Essentially, the weight value, which reflects physical, chemical and biological properties, can be assigned to vertices (atom centers), edges (bonds), or higher order simplexes (cluster of atoms), depending on the biomolecular structure, function, and dynamics properties\cite{meng2019weighted}. In this way, weighted persistent homology can be characterized into three major categories, i.e., vertex-weighted\cite{edelsbrunner1992weighted,bell2017weighted,guibas2013witnessed,buchet2016efficient,KLXia:2015c,xia2015multiresolution}, edge-weighted\cite{petri2013topological,Binchi:2014jholes,KLXia:2015c,KLXia:2014persistent,cang:2017topologynet,cang:2018representability}, and simplex-weighted models\cite{dawson1990homology,ren2018weighted,wu2018weighted}. Among them, the localized (weighted) persistent homology (LPH) and interactive persistent homology (IPH) are found to be of great importance in the classification and clustering of DNA structures and trajectories\cite{meng2019weighted}, and protein ligand interactions\cite{cang:2018representability}.

In this paper, for the first time, we apply the localized persistent homology and interactive persistent homology in the study of osmolyte molecular aggregation and their hydrogen-bonding networks. Similar to our previous works, we use persistent Betti number (PBN) and persistent entropy (PE) to quantitatively characterize the topological features from LPH and IPH. More specifically, using LPH, we have revealed that TMAO and urea have very different local topologies. Local network structures are observed in TMAO systems. With the concentration increase, the circles within these networks show a huge increase in their total numbers and a sharp decrease in their relative sizes. In contrast, urea shows two distinguishable local topological features, i.e., local clusters around 6 \AA~ and global-scale circle structures at around 12 \AA. Further, we have demonstrated that our global-scale IPH based persistent radial distribution function (PRDF) is similar to the traditional radial distribution function (RDF) and can be used to characterize the double layer information. Moreover, a new local-scale  PRDFs can be generated from our local-scale IPH model. Essentially, in global-scale IPH, each osmolyte molecule interacts with all the water molecules in the systems. However, in global-scale IPH, water molecules are classified into different cells based on the Voronoi diagram of osmolyte molecules. The interaction happens between the osmolyte molecule and the water molecules within its Voronoi cell and between two closest Voronoi cells. This classification is naturally embedded in the filtration process of IPH analysis. Further, IPH based PBNs and PEs can also be used to study the interaction patterns between osmolyte molecules and water molecules. Other than the osmolyte systems, our weighted persistent homology models can also be used to analyze networks, graphs, biomolecules, etc.

The paper is organized as follows. A brief introduction of persistent homology and two weighted persistent homology models are given in Section \ref{sec:theory}.
The main results are presented in Section \ref{sec:results}. The LPH based molecular aggregation and hydrogen-bonding networks is discussed in Section \ref{sec:LPH}. The IPH based topological features for osmolyte-water interaction networks are discussed in Section \ref{sec:IPH_results}. The paper ends with a conclusion.

\section{Theory and models}\label{sec:theory}
In this section, we will give a brief introduction of persistent homology and weighted persistent homology. Three types of persistent functions, including persistent Betti number, persistent entropy and persistent radial base function, will be discussed in details. A general description of the two WPH models, i.e., localized persistent homology and interactive persistent homology, will also be presented.

\subsection{Persistent homology}
Persistent homology has been proposed to study the ``shape" of data. It has became more and more popular in a variety of fields, including shape recognition \cite{DiFabio:2011}, network structure \cite{Silva:2005,LeeH:2012,Horak:2009}, image analysis \cite{Carlsson:2008,Pachauri:2011,Singh:2008,Bendich:2010,Frosini:2013}, data analysis \cite{Carlsson:2009,Niyogi:2011,BeiWang:2011,Rieck:2012,XuLiu:2012}, chaotic dynamics verification \cite{Mischaikow:1999}, computer vision \cite{Singh:2008}, computational biology \cite{Kasson:2007,YaoY:2009, Gameiro:2013}, amorphous material structures\cite{hiraoka:2016hierarchical,saadatfar:2017pore}, etc. Researchers have developed many elegant softwares, including JavaPlex \cite{javaPlex}, Perseus  \cite{Perseus}, Dipha \cite{Dipha}, Dionysus \cite{Dionysus}, jHoles \cite{Binchi:2014jholes}, GUDHI\cite{gudhi:FilteredComplexes}, etc\cite{fasy:2014introduction}. Visualization methods, including persistent diagram\cite{Mischaikow:2013}, persistent barcode\cite{Ghrist:2008barcodes}, and persistent landscape\cite{Bubenik:2007,bubenik:2015}, have also been proposed. In this section, we only present a brief introduction of persistent homology. A more detailed description of its mathematical background can be found in references\cite{Edelsbrunner:2002,CZCG05,Zomorodian:2005}.

%A more detailed description of its mathematical background is given in the appendix and can also be found in the references\cite{Edelsbrunner:2002,CZCG05,Zomorodian:2005}.

Persistent homology studies the topological invariant called Betti numbers, which include $\beta_0$, $\beta_1$, $\beta_2$, etc. Geometrically, $\beta_0$ represents the number of connected components, $\beta_1$ represents the number circles, rings or loops, and $\beta_2$ represents the number of voids or cavities. Persistent homology enables a multiscale representation of topological invariants from simplicial complexes. Generally speaking, a network or graph is a special kind of simplicial complex with only 0-simplexes (nodes or vertexs) and 1-simplexes (edges). Simplicial complexes can have higher-dimensional components\cite{giusti2016two}, such as 2-simplexes and 3-simplexes, which can be geometrically viewed as triangles and tetrahedrons, respectively. Moreover, the key concept in persistent homology is the filtration. For example, if we have a point cloud data, we can associate each point with an identical-sized sphere and assign its radius as the filtration parameter. With the increase of filtration value, these spheres will systematically enlarge and further overlap with each other to form simplexes. Roughly speaking, an edge between two points is formed when the two corresponding spheres overlap. A triangle is formed when each of two spheres (of the three corresponding spheres from triangle vertices) overlap. And a tetrahedron is formed when each three spheres (of the four corresponding spheres from tetrahedron vertices) overlap. At each filtration value, all the simplexes, i.e., vertices, edges, triangles, tetrahedrons, form a simplicial complex. From it, topological invariants, i.e., Betti numbers, can be calculated.  Through a systematical variation of a filtration parameter, a series of simplicial complexes from different scales are generated. Some topological invariants persistent longer in these simplicial complexes, while others quickly disappear when the filtration value changes. In this way, the ``lifespan" of the topological invariants (circles, loops, etc) in these simplicial complexes provides a natural geometric measurement. More specifically, the lifespan, known as the persistence, measures how ``large" are the circles, loops and voids in the system.

We call a filtration value, at which a topological invariant appears or disappears, birth time (BT) and  death time (DT), respectively. In this way, each topological invariant has a ``lifespan" defined by its birth and death time. Essentially, the lifespan provides a geometric measurement of the topological invariant. If we use a one-dimensional bar, started at BT and ended at DT, to represent each homology generator, a barcode representation is generated. We can denote barcodes as follows,
\begin{eqnarray}
\{ L_{k,j}=[a_{k,j}, b_{k,j}] | k=0,1,...; j=1,2,3,....,N_k \},
\end{eqnarray}
where parameter $k$ represents the $k$-th dimension. For data points located in Euclidean space, normally we only consider $k=0,1,2$. Parameter $j$ indicates the $j$-th topological invariant and $N_k$ is the number of $\beta_k$ topological invariant. The $a_{k,j}$ and $b_{k,j}$ are BTs and DTs. For simplification, we can define the set of $k$-th dimensional barcodes as,
$$ L_{k}= \{ L_{k,j}, j=1,2,3,....,N_k\}, \quad k=0, 1, ....$$
Various functions can be defined on the barcodes.

\paragraph{Persistent Betti number}
Based on the persistent homology results, different functions are proposed to represent or analyze the topological information\cite{Edelsbrunner:2002,Carlsson:2009,bubenik:2015,Chintakunta:2015}. Among them is the persistent Betti number (PBN), which is defined as the summation of all the $k$-th dimensional barcodes,
\begin{eqnarray}\label{eq:PBN}
f(x;L_{k})= \sum_{j} \chi_{[a_{k,j},b_{k,j}]}(x), \quad k=0, 1, ....
\end{eqnarray}
Function $\chi_{[a_{k,j}, b_{k,j}]}(x)$ is a step function, which equals to one in the region $[a_{k,j}, b_{k,j}]$ and zero elsewhere. This equation transforms the $k$-dimensional barcodes into a one-dimensional function. We can also define an average PBN as follows,
\begin{eqnarray}\label{eq:sPBN}
f(x;L_{k})= \frac{1}{N}\sum_{j} \chi_{[a_{k,j},b_{k,j}]}(x), \quad k=0, 1, ....
\end{eqnarray}
Note that $N$ can be the total number of atoms, $\beta_k$ bars ($N_k$), or other measurements.

\paragraph{Persistent entropy}
Persistent entropy (PE) has been proposed\cite{Merelli:2015topological,Chintakunta:2015,Rucco:2016,Xia:2018multiscale} to measure the system disorder. For the $k$-th dimensional barcodes, it is defined as,
\begin{eqnarray}\label{eq:filtrationM}
S_k=\sum_j^{N_k} - p_{k,j} \ln(p_{k,j}), \quad k=0, 1, ...,
\end{eqnarray}
with the probability function,
\begin{eqnarray}\label{eq:pi}
p_{k,j}=\frac{b_{k,j}-a_{k,j}}{\sum_j (b_{k,j}-a_{k,j})}, \quad k=0, 1, ....
\end{eqnarray}
The expression of PE can be simplified as follows,
\begin{eqnarray}\label{eq:MPE}\nonumber
S_k=\ln\left(\sum_j^{N_k} (b_{k,j}-a_{k,j})\right)- \frac{\sum_j^{N_k}\left( (b_{k,j}-a_{k,j}) \ln(b_{k,j}-a_{k,j}) \right)}{\sum_j^{N_k} (b_{k,j}-a_{k,j})}, \\ \quad k=0, 1, ....
\end{eqnarray}

It should be noticed that PE is different from the general entropy used in molecular dynamic simulation. Generally speaking, PE characterizes the ``topological regularity". For instance, crystals have very consistent three-dimensional structures. They will generate highly regular barcodes thus a large PE value. Topologically, a large PE value means that the structure is more regular and ``lattice-like", while a lower PE value means the components in the structure are randomly distributed with no consistent patterns.

\paragraph{Persistent radial distribution function}
Based on the $\beta_0$ barcode, we propose the persistent radial distribution function (PRDF) as follows,
\begin{eqnarray}\label{eq:PRDF}
f(x;L_{0})= \frac{x_t}{N_0}\sum_{j}\frac{\delta(x-b_{0,j})}{4\pi x^2}.
\end{eqnarray}
Here $x_t$ is the filtration value when the PBN reduces to one, i.e., only one connected component. Integer $N_0$ is the total number of $\beta_0$ bars. Essentially, if we consider the global interactive persistent homology, our PRDF will result in the common RDF\cite{chandler1987introduction}. If we use the local interactive persistent homology, our PRDF will focus on the interaction within each cell of the Voronoi diagram. Detailed discussion is given in Section \ref{sec:InteractivePH}.

\subsection{Weighted persistent homology}
Weighted persistent homology models have been proposed to incorporate physical, chemical and biological properties into topological modeling\cite{meng2019weighted}. They can also be designed to characterize local topological information and certain special interaction patterns. In this paper, we will focus on two WPH models, i.e., localized persistent homology and interactive persistent homology.

\subsubsection{Localized persistent homology}
\begin{figure}
	\begin{center}
		\captionsetup[subfigure]{labelformat=empty,labelsep=none}
		\includegraphics[scale = 0.3]{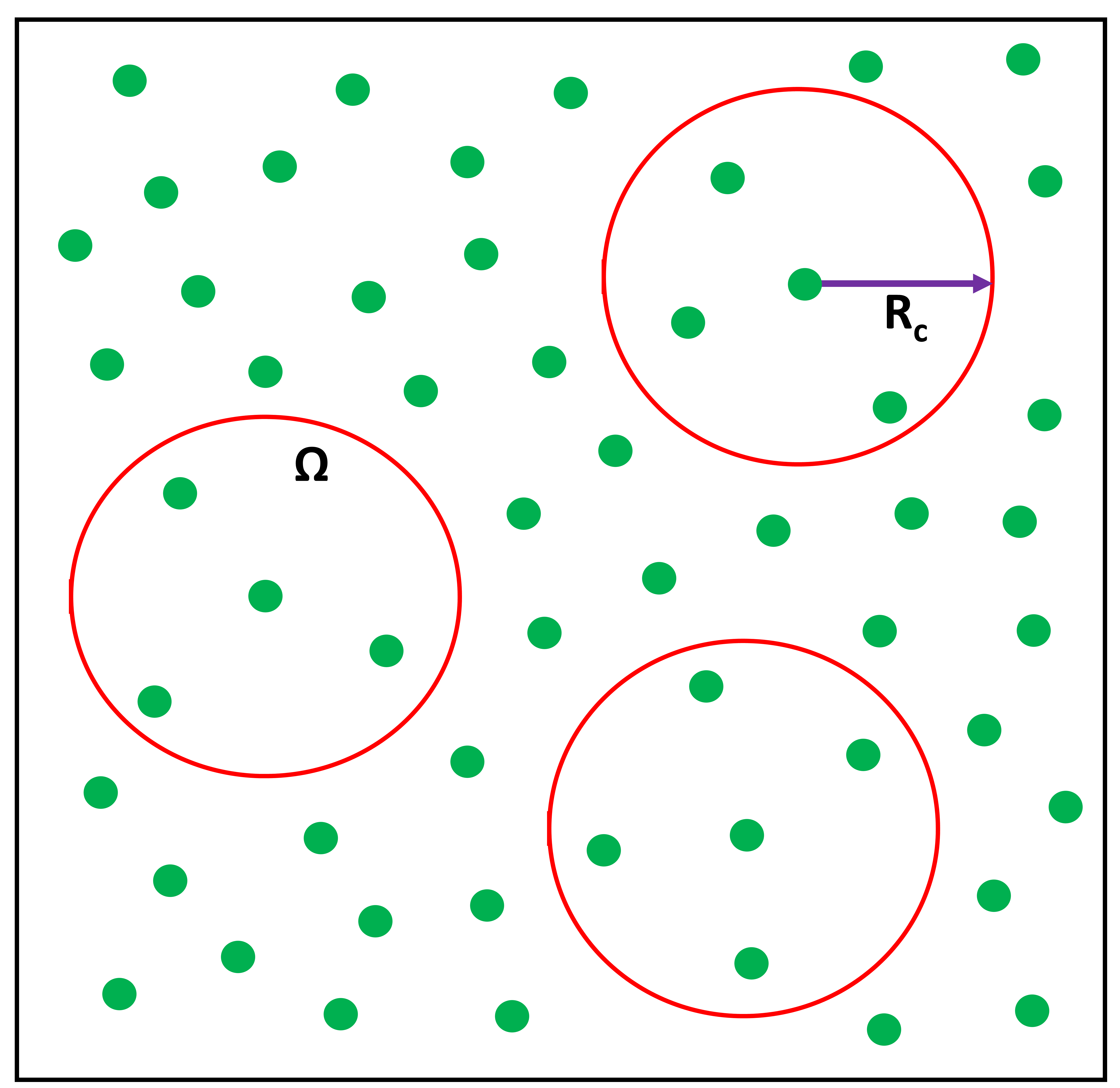}
		\caption{A schematic representation of the localized persistent homology (LPH) analysis. The red circle (a sphere in 3D) specifies the local region. Persistent homology is carried out on all the molecules within the local region.}
		\label{fig:Local_PH}
	\end{center}
\end{figure}

The design of our LPH model is greatly inspired by the great success of element specific persistent homology (ESPH)\cite{cang:2017topologynet,cang:2017integration}. Different from all previous topological models, which consider the data/structure as an inseparable system, ESPH decomposes the data/structure into a series of subsets made of certain type(s) of atoms, which have been found to characterize very well various biological properties, such as hydrophobic or hydrophilic interactions \cite{cang:2017topologynet,cang:2017integration,nguyen:2017rigidity,cang:2017analysis,cang:2018representability,wu:2018quantitative, nguyen2018mathematical}. Moreover, our LPH model is very different from traditional persistent local homology (PLH)\cite{bendich2007inferring,bendich2012local,ahmed2014local,bendich2015multi,fasy2016exploring,munkres2018elements}. Mathematically, PLH model studies the relative homology groups between a topological space and its subspace, while LPH explores the homology groups from local topology. Previously, LPH was used to characterize local topological features of biomolecular structure or complexes\cite{meng2019weighted}. In LPH, the structure is decomposed into a series of local domains or regions, that may overlap with each other, and persistent homology analysis is then systematically applied on part (or all) of these local domains ore regions. In this paper, our main focus is to characterize the local features, such as ion clustering, double layer, local aggregations, etc, that are widely existed in ion or molecular aggregation and hydrogen-network systems.

Mathematically, the global persistent homology analysis studies the whole system, while the localized persistent homology is performed on a local region, domain or subspace. In the current paper, we define the subspace as a sphere with radius ($R_c$). More specifically, a sphere of radius $R_c$ is considered around each molecule (either osmolyte or water molecule) and only the molecules within this sphere are chosen for the localized persistent homology analysis. Figure \ref{fig:Local_PH} shows a schematic configuration in which a particular molecule is selected and a sphere of radius $R_c$ is drawn around it. The molecules within this enclosure are chosen as the local neighbors of the central molecule. The persistent homology analysis is carried out for the selected molecules to generate the local persistent barcodes. This procedure is repeated for each molecule in the configuration. In other words, each molecule in a given configuration is associated with its local neighbors which dictate its local structure.

\subsubsection{Interactive persistent homology}\label{sec:InteractivePH}

\begin{figure}
	\begin{center}
		\captionsetup[subfigure]{labelformat=empty,labelsep=none}
		\includegraphics[scale = 0.6]{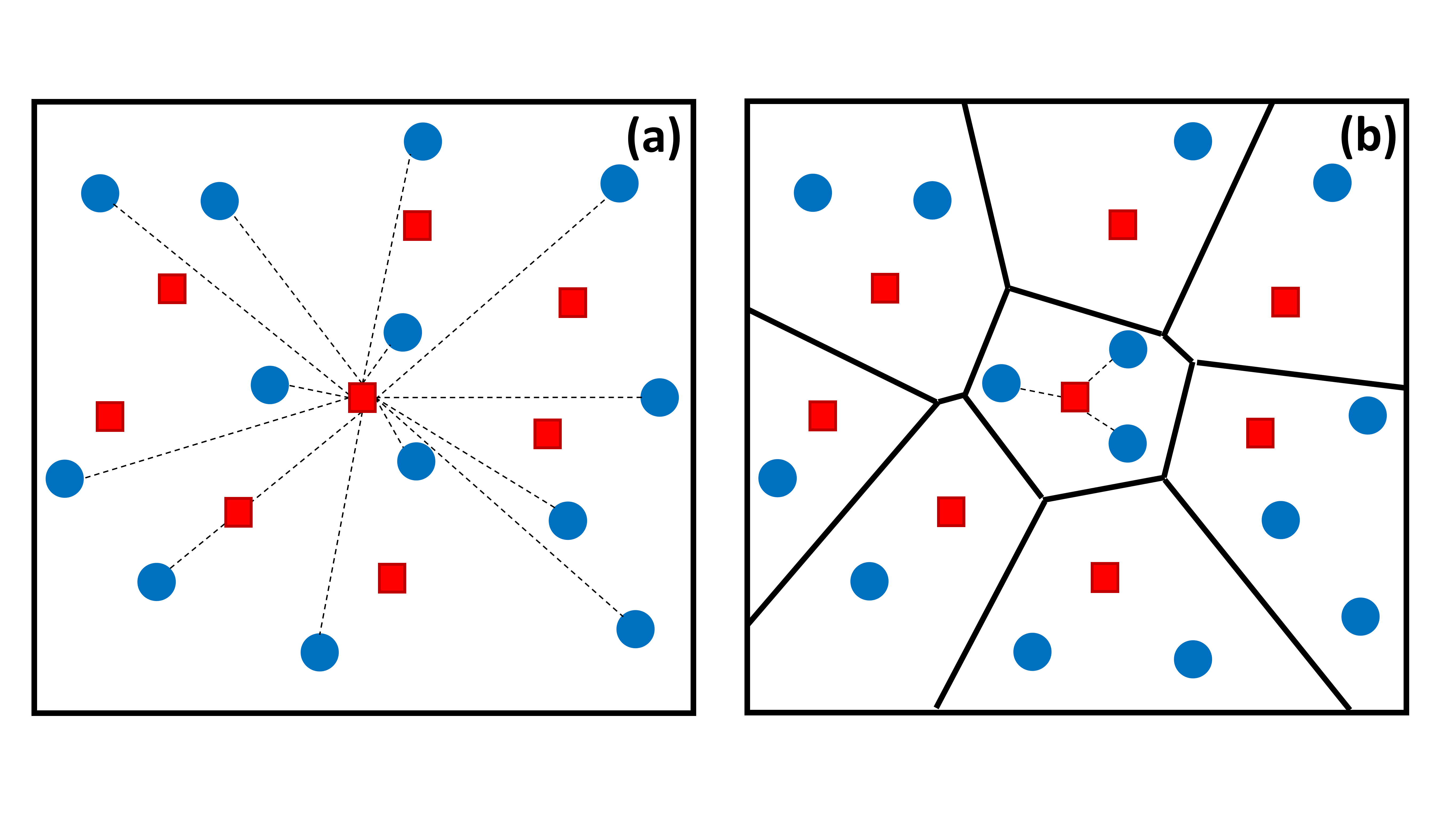}
		\caption{Illustration of interactions in global-scale and local-scale interactive persistent homology (IPH). In global-scale IPH model, an osmolyte molecule (red rectangle) interacts with all the water molecules (blue dots). In local-scale model, an osmolyte molecule (red rectangle) interact only with water molecules (blue dots) in its Voronoi cell, and Voronoi cell-cell interactions happen only between closest neighbours. }
		\label{fig:Global_Local_RDF}
	\end{center}
\end{figure}

The interactive persistent homology (IPH) is proposed to study the interaction between proteins and ligands\cite{cang:2018representability}. The essential idea is to study the topological invariant of the interaction networks, which are formed between protein atoms and ligand atoms. More specifically, for a protein-ligand complex, an interaction matrix can be built with its element as the Euclidean distance between two atoms. However, if two atoms come from the some molecule (either protein or ligand), its distance is set to infinity, meaning they will never interact in IPH. In this way, the IPH model can be used in the characterization of the protein-ligand interactions. Actually, IPH based machine learning models are found to deliver the best results in protein-ligand binding affinity prediction\cite{cang:2017analysis,cang:2018representability,nguyen2018mathematical}.

In this section, we use IPH models to characterize the interactions between osmolyte and water molecules. Two different models, i.e., global-scale IPH and local-scale IPH, are considered. In global-scale model, when an osmolyte molecule is selected and the distances ($d_{ij}$) between all water molecules to this osmolyte molecule are considered. More specifically, suppose there are $N_{wtr}$ number of water molecules, a global-scale IPH matrix of size $(N_{wtr}+1)\times(N_{wtr}+1)$ can be constructed between a selected osmolyte molecule and all water molecules as follows,
\begin{eqnarray}\label{fig:IPH}
M_{ij} &=& \left\{
\begin{array}{ll}
d_{ij}, &  {\rm if}~ T_{ype}(i) \neq T_{ype}(j);  \\
\infty, & {\rm otherwise}  \\
\end{array}
\right.
\end{eqnarray}
Here $T_{ype}(i)$ is used to tell if the $i$-th molecule is osmolyte or water, i.e., type of the molecule. If there are $N_{sml}$ number of osmolyte molecule, we can construct a total $N_{sml}$ number of gloabl-scale IPH matrixes. From these matrixes, PRDF as in Eq. (\ref{eq:PRDF}) can be calculated and the average of these PRDFs will characterize the same physical properties as the traditional radial distribution function\cite{chandler1987introduction}.

In local-scale IPH, a similar IPH matrix as in Eq.(\ref{fig:IPH}) is considered. But this new IPH matrix is now of size $(N_{wtr}+N_{sml})\times(N_{wtr}+N_{sml})$, meaning all distances between water and osmolyte molecules are considered simultaneously. The new IPH matrix based filtration characterizes dramatically different topological information. More specifically, molecules with shorter distances to their neighbors will form connections at earlier stage of the filtration. In this way, a Voronoi diagram will naturally form when water molecules connect to their center osmolyte molecule. Later, Voronoi cells will merge with closest neighbors to become a well-connected entity. The $\beta_0$ barcodes capture very well the above topological information. And the corresponding PRDFs describe the local interactions within the Voronoi cells.

A comparison between global-scale and local-scale IPH is illustrated in Figure \ref{fig:Global_Local_RDF}. Essentially, each osmolyte molecule can interact directly with all water molecules in global-scale model and the resulting PRDF (from $\beta_0$ barcodes) characterizes the same physical properties as RDF. In local-scale IPH model, only the interactions between the osmolyte molecule and water molecules in its Voronoi cell, and the Voronoi cell-cell interactions are captured in $\beta_0$ bars. It should be noticed that the corresponding PRDFs only describe the local interaction information, and they are very different from the traditional RDF.

\section{Results and discussions}\label{sec:results}

We use the weighted persistent homology models to investigate the local topological structure of two osmolytes, namely, trimethylamin N-oxide (TMAO) and urea. Two models, i.e., localized persistent homology and interactive persistent homology, are used to reveal the local topological features in the aggregation, hydrogen networks, and their interactions.

\subsection{MD simulation and data generation}

We consider the same molecular dynamics (MD) setting as in the paper \cite{xia2019persistent}. More specifically, we consider GROMACS-5.1.2 \cite{abraham2015gromacs,berendsen1995gromacs} for the MD simulation. The four point (TIP4P-EW) \cite{horn2004development} water model are used, and Kast model \cite{kast2003binary} is adopted for TMAO whereas the urea model is from AMBER package\cite{pearlman1995amber}. Two osmolytes with eight different concentrations, from 1M to 8M, in pure water are studied, respectively. To construct the initial state, urea/TMAO molecules are randomly distributed using insert-molecules utility in GROMACS, after that 3000 water molecules are inserted randomly into the cubic simulation box. We carry out the equilibration process under NVT conditions (Temperature = 300 K) for 10 ps and then under NPT conditions for 100 ps using 2 fs time step, Berendsen thermostat $\tau$ = 0.1ps) and barostat ($\tau$= 2 ps). LINCS algorithm \cite{hess1997lincs} is used for bonds and the angles constriction.  Further, we carry out three repeats under NPT conditions for 100 ns with Berendsen thermostat (Temperature = 300 K, $\tau$ = 0.1 ps)  Parrinello-Rahman barostat (Pressure = 1 bar, $\tau$ = 2 ps) and using a time step of 2 fs. The integrate Newton's equation of motion is done by using a leap-frog algorithm. A cut-off of 1.0 nm is used for both van der Waals (VDW) interaction and short-range electrostatic interaction. Particle mesh Ewald (PME) \cite{essmann1995smooth} method is employed to deal with the long-range electrostatic interactions. The configuration trajectories are output every 1 ps.

\subsection{LPH for molecular aggregation and hydrogen-bonding networks}\label{sec:LPH}

\begin{figure}
	\begin{center}
		\captionsetup[subfigure]{labelformat=empty,labelsep=none}
		\includegraphics[scale = 0.35]{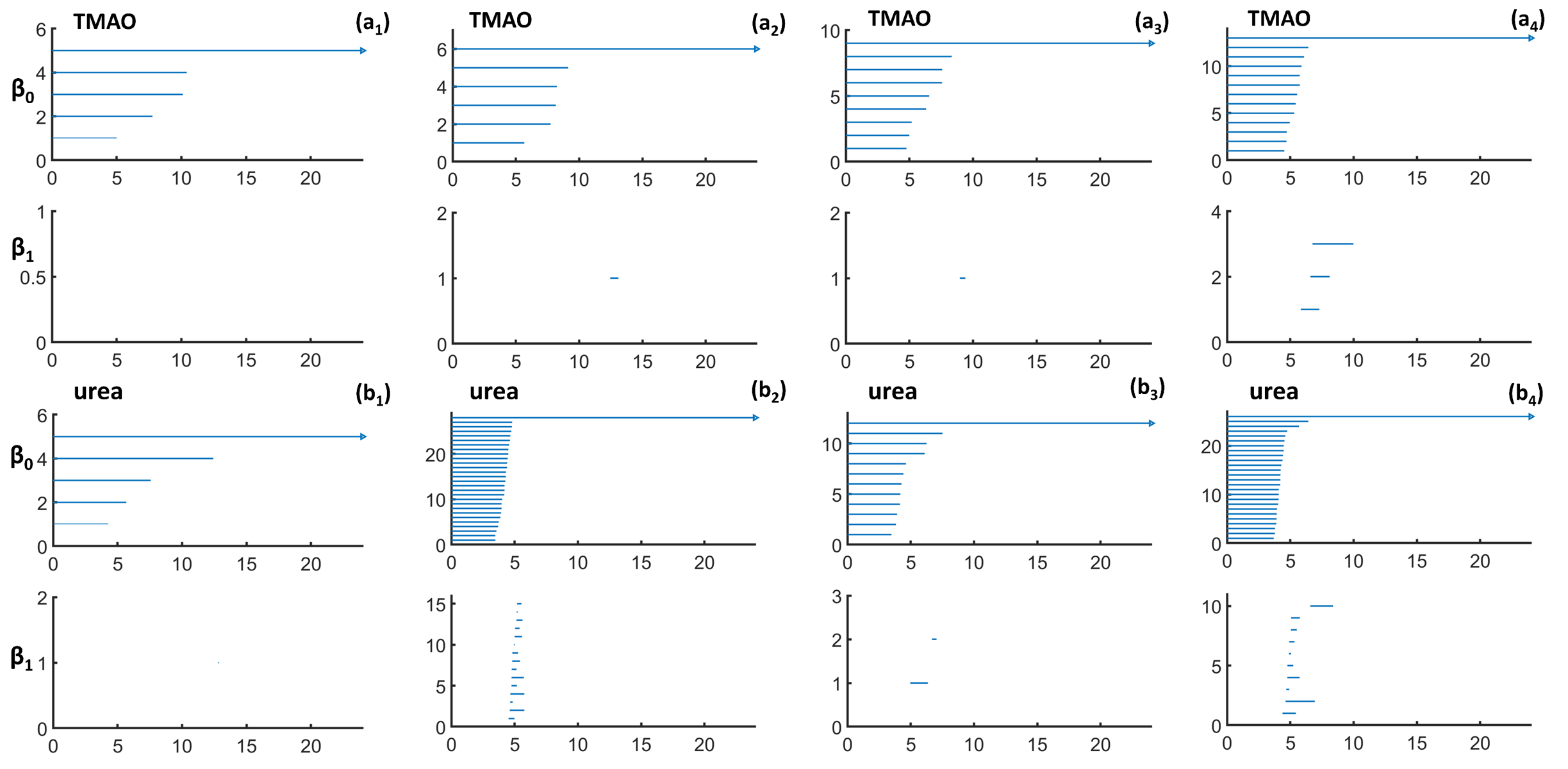}
		\caption{The local persistent barcodes for TMAO and urea aggregation. TMAO and urea molecules are coarse-grained as their nitrogen and carbon atoms. The barcodes are generated from a randomly picked molecule at the last frame of the simulation. A cutoff radius of 9\AA~ is used. Four different concentrations, i.e., 2M, 4M, 6M, and 8M are considered. Roughly speaking, both $\beta_0$ and $\beta_1$ bars tend to increase with the concentration.}
		\label{fig:Barcodes}
	\end{center}
\end{figure}

In this section, localized persistent homology is used to explore the topological fingerprint of molecular aggregation and hydrogen-bonding network at a local scale. We systematically study the osmolyte and water molecular aggregation patterns within a certain local domain, specified by a sphere with a radius from 7\AA~ to 15 \AA. The corresponding PBNs and PEs are used to quantitatively characterize the intrinsic local topology information.

Computationally, in TMAO systems, there are 63, 125, 204, 290, 400, 533, 700 and 887 TMAO molecules with 3000 water molecules from concentration 1M to 8M, respectively. In urea systems, there are 60, 120, 192, 267, 352, 450, 555 and 681 urea molecules with 3000 water molecules respectively. To analyze the topology in molecular aggregation and hydrogen-bonding networks, the TMAO and urea molecules are coarse-grained as their nitrogen and carbon atoms, respectively. The water molecules are coarse-grained as their oxygen atoms. Since the configuration data is obtained from an NPT simulation, the size of simulation box is allowed to adjust for each configuration to attain equilibrium conditions. Periodic boundary condition is used in the specification of local domains. For each simulation, we consider 101 frames (or configurations) sampled at equal intervals from the simulation trajectory. Our topological analysis is performed on these 101 frames.

\subsubsection{LPH based topological features of two types of osmolytes}
\begin{figure}
	\begin{center}
		\captionsetup[subfigure]{labelformat=empty,labelsep=none}
		\includegraphics[scale = 0.3]{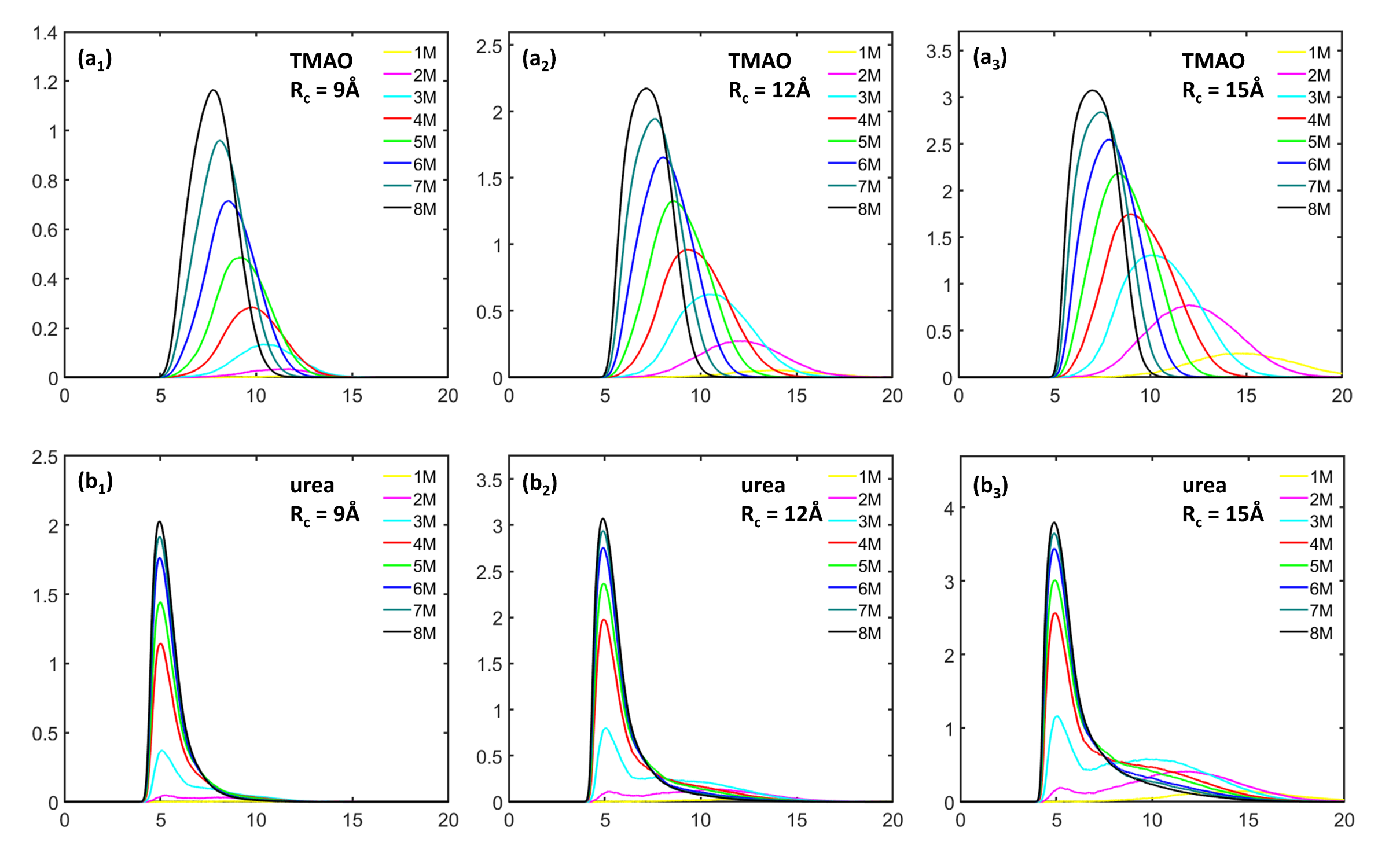}
		\caption{The comparison of average $\beta_1$ PBNs for TMAO and urea using three different cutoff radii namely., $R_c = 9${\AA}, $12${\AA} and $15${\AA} respectively. The $\beta_1$ PBNs are averaged over all the frames and all molecules in each frame. It can be seen that, TMAO and urea show dramatically different local topological characteristics.}
		\label{fig:Topo_PBN_sml}
	\end{center}
\end{figure}

To facilitate an intuitive understanding of local topological information of molecular aggregation, we demonstrate the persistent barcodes calculated from TMAO and urea systems with a cutoff radius of $R_c = 9${\AA}. More specifically, we consider the last configurations of the MD simulations from four different concentrations. An osmolyte molecule is randomly chosen from these last frames and its neighbouring osmolyte molecules located within the cutoff radius $R_c = 9${\AA}~ are also selected. Persistent homology analysis is then applied on these selected molecules. The results from TMAO and urea systems are demonstrated in Figures \ref{fig:Barcodes}. In both TMAO and urea, the total number of $\beta_0$ bars roughly increase with concentration (M), indicating the aggregation of neighboring molecules with the concentration. The $\beta_1$ bars also seem to appear more and more frequently with the increase of the concentration.

All the above results are based on a randomly picked molecule in the last frames of the simulation trajectories and can not characterize the overall behavior very well. To have a better comparison, we consider the ensemble average. More specifically, for each frame, the local persistent barcode from each molecule is calculated and then averaged. It should be noticed that we use the periodic boundary condition to include all ``neighboring" molecules. This process is repeated for all 101 frames in each trajectory. We represent each persistent barcode as their PBN and PE. These PBNs are then averaged over all the frames and all the molecules in each frames to generate a single PBN for each simulation or trajectory. The PEs are averaged over all the molecules in each frame, so that a total 101 PEs are obtained form each simulation.

\begin{figure}
	\begin{center}
		\captionsetup[subfigure]{labelformat=empty,labelsep=none}
		\includegraphics[scale = 0.3]{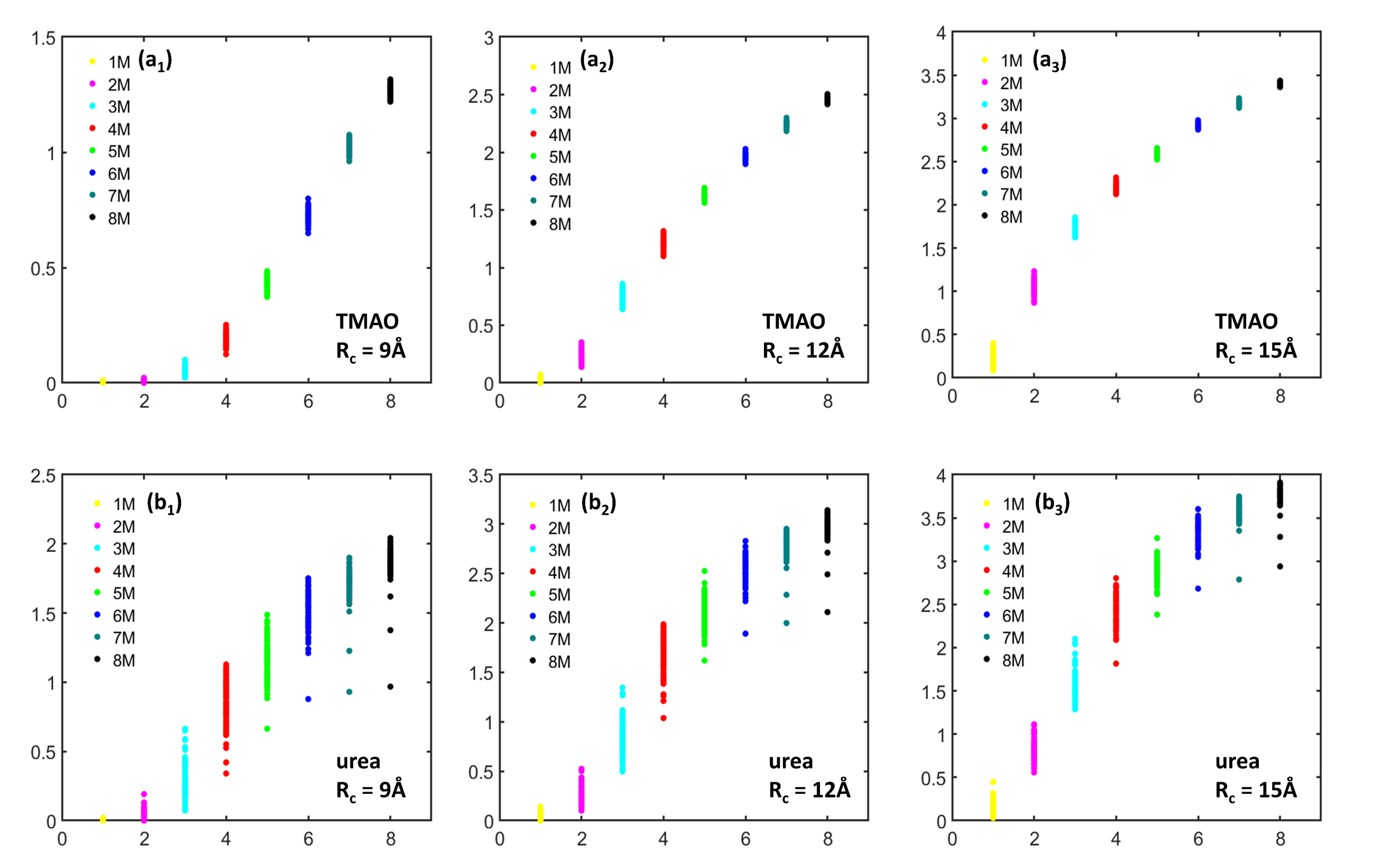}
		\caption{The comparison of average persistent entropies for TMAO and urea using three different cutoff radii, i.e., $R_c = 9${\AA}, $12${\AA} and $15${\AA}. The PEs are averaged over all the molecules in each frames, thus a total 101 PEs are obtained for each simulation. It can be seen that, for a small cutoff radius of $R_c=9${\AA}, both TMAO and urea PEs at 1M are almost all zero. Further, the average PEs for both systems increase with the concentration, but their PE variances have very different properties. The TMAO PE variance decreases with concentration while urea PE variance increases.}
		\label{fig:Topo_PE_sml}
	\end{center}
\end{figure}

Figure \ref{fig:Topo_PBN_sml} shows the $\beta_1$ PBNs obtained for the TMAO  and urea system at eight different concentrations, under three different cutoff radii, i.e., $R_c = 9${\AA}, $12${\AA} and $15${\AA}, respectively. As stated above, $\beta_1$ bars represent the ring, circle and loop structures in the systems. For TMAO systems, at each cutoff radius, the peak value of the local $\beta_1$ PBNs systematically increases with the concentration, indicating more and more circle structures are generated. At the same time, the position of these peak values shifts from around 13\AA~ to 7\AA, meaning that the size of circles systematically decreases. When we consider larger cutoff radii, similar topological patterns are observed. However, PBNs from lower concentration systems increase much faster, even though all PBNs increase with the cutoff radius. This results indicate that for lower concentration systems, there exists large-sized topological features that can not be well characterized by LPH with a small cutoff radius. For urea systems, their PBNs have a dramatically different behavior in comparison with TMAO. Roughly speaking, there are two types of peak for urea systems, especially urea systems at lower concentrations. One type of peak is located around 5 \AA, and the other is around 10\AA~ to 12\AA. The peak at 5\AA~ appears even at very low concentrations and its magnitude keeps increasing with the concentration rise. The shape of this peak is much narrower that of those of TMAO PBNs. The second type of peak can only be well observed at lower concentrations. It has much smaller magnitude compared with that of the first type of peak.

From Figure \ref{fig:Topo_PBN_sml}, we can see that TMAO and urea demonstrate dramatically different local topological characteristics. Essentially, TMAO shows a regular local network structure. The size and total number of the circle structures from these networks consistently decrease and increase with the concentration, respectively. In contrast, urea shows a cluster-like local aggregations. Urea molecules form local clusters, whose size stays relative consistent but the total number consistently increases with concentration. Compared with our results from persistent homology analysis of the whole system \cite{xia2019persistent}, LPH focuses more on the local topological information and systematically attenuates the influence from global topological features.

Other than PBNs, we can also calculate PEs from the LPH barcodes and use them to characterize the ``topological regularity". Figure \ref{fig:Topo_PE_sml} demonstrates the $\beta_1$ PEs for both TMAO and urea at eight concentrations and three cutoff radii as stated above. Note that for each simulation or trajectory, we consider 101 configurations or frames and generate 101 $\beta_1$ PEs. It can be seen that, at a small cutoff radius, the PE values from 1 M concentration system is almost all zeros, meaning that there is almost no circle structures at local scale. This is consistent with the PBN profile in Figure \ref{fig:Topo_PBN_sml}. Further, the average PE value increases systematically with the concentration for both TMAO and urea. However, the PE variance shows a very different behavior. With the concentration increase, the TMAO PE variance systematically decreases, while urea PE variance consistently increases. These results are also consistent with our findings from persistent homology analysis of the whole system\cite{xia2019persistent}. Essentially, with the concentration increase, all osmolyte systems become topological more and more disordered. However, the variation of topological regularity for each trajectory decreases in TMAO systems but increases in urea systems.

\subsubsection{LPH based topological features of hydrogen-bonding networks}

\begin{figure}
	\begin{center}
		\captionsetup[subfigure]{labelformat=empty,labelsep=none}
		\includegraphics[scale = 0.22]{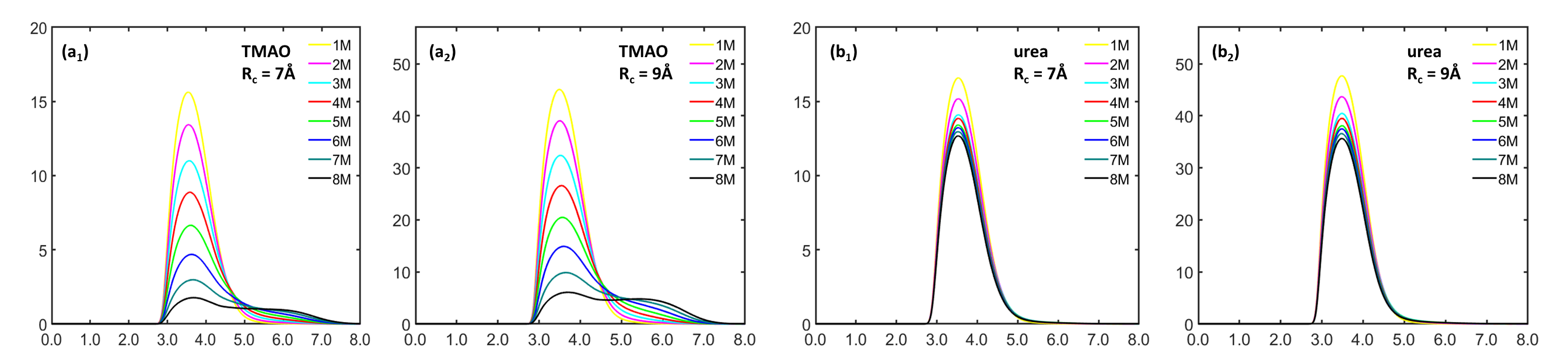}
		\caption{The comparison of average $\beta_1$ PBNs for hydrogen bonding networks from TMAO and urea using two different cutoff radii, i.e., $R_c = 7${\AA} and $R_c = 9${\AA}. The coarse-grained representation of water as its oxygen atom is considered. The PBNs are averaged over all the molecules and configuration numbers. It can be seen that, TMAO and urea show very different topological characteristics.  }
		\label{fig:Topo_PBN_wtr}
	\end{center}
\end{figure}

In our hydrogen-bonding network analysis, we only consider the topological features for water molecules at a local scale. Similar to osmolyate systems, LPH analysis is carried out for each water molecule together with its neighbours located within a cutoff radius $R_c$. For each frame, we systematically go over all the 3000 water molecules and calculate 3000 local persistent barcodes. Again periodic boundary condition is considered to include all ``neighboring" water molecules. The process is repeated over all the 101 frames in each trajectory. A single $\beta_1$ PBN is generated for each simulation by averaging $\beta_1$ PBNs over all the 101 frames and all the 3000 water molecules in each frame. The $\beta_1$ PEs are averaged over the 3000 water molecules in each frame, so that a total 101 $\beta_1$ PEs are obtained from each simulation. Two cutoff radii, i.e, $R_c = 7${\AA} and $R_c = 9$ {\AA}, are considered in our LPH analysis.

Figure \ref{fig:Topo_PBN_wtr} shows the comparison of average $\beta_1$ PBNs for TMAO and urea hydrogen-bonding networks. For TMAO systems, it can be seen that their PBNs have a peak value located at around 3.5\AA. With the concentration increase, the peak value of TMAO PBNs gradually decreases. In the meantime, there is a consistent rise of the PBN values in the range from around 4.5 \AA~ to 7.0 \AA. Even though all PBNs significantly increase with the cutoff radius, the general PBN profile pattern from eight different concentrations is highly consistent. Similar to TMAO, urea PBNs also have a peak value at filtration value 3.5\AA. The peak value slightly decreases with the concentration increase. Further, the general PBN profile pattern from eight different concentrations shares a remarkable similarity at two different local scales, even though the PBN peak values are systematically increased.

From Figure \ref{fig:Topo_PBN_wtr}, we can see that TMAO and urea hydrogen-bonding networks demonstrate dramatically different local topological characteristics. For TMAO hydrogen-bonding networks, with the concentration increase, there is a systematical decrease of small-sized circle structures as well as an increase of relatively large-sized circle structures. For urea hydrogen-bonding networks, there is only a slight decrease of small-sized circle structures and no significantly increase of large-sized circle structures. More interestingly, if we compare the our LPH results with the ones from the whole hydrogen-bonding network in both ion and osymolate systems\cite{xia2018persistent,xia2019persistent}, we can see that there exists a great similarity in their PBNs. Essentially, TMAO and urea hydrogen-bonding networks show two types of topological behaviors. With the concentration increase, TMAO molecules tend to destroy the local hydrogen-bonding networks, resulting a significantly increase of large circle structures. In contrast, the urea molecules have a much less impact on the hydrogen-bonding networks.

\begin{figure}
	\begin{center}
		\captionsetup[subfigure]{labelformat=empty,labelsep=none}
		\includegraphics[scale = 0.22]{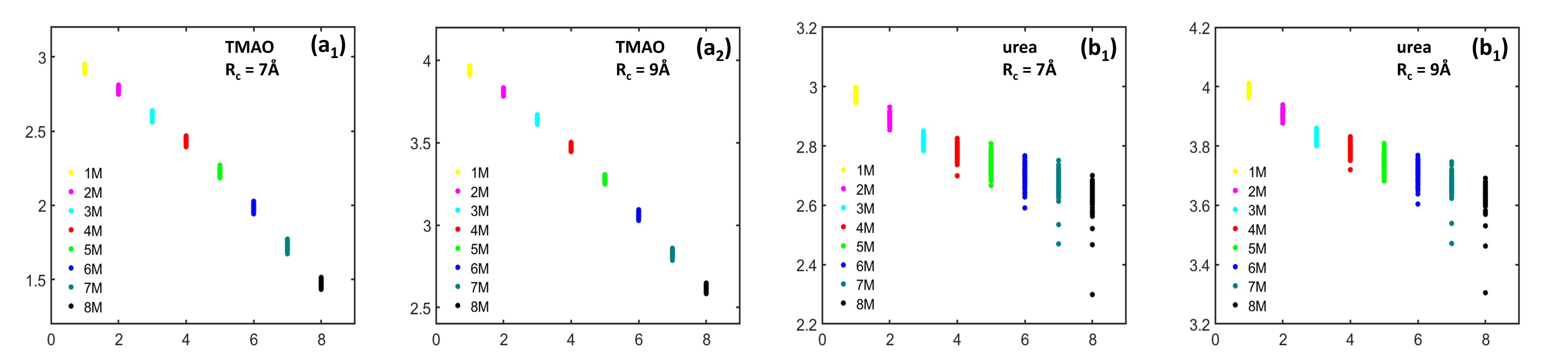}
		\caption{The comparison of average $\beta_1$ PEs for hydrogen-bonding networks from TMAO and urea systems using two different cutoff radii, i.e., $R_c = 7${\AA} and $R_c = 9${\AA}. The PEs are averaged over all the water molecules in each frames, thus a total 101 PEs are obtained for each simulation. It can be seen that, the average PE decreases with the concentration for both TMAO and urea. However, the PE variance for urea systematically increases.}
		\label{fig:PE_Water}
	\end{center}
\end{figure}

Persistent entropy from the LPH barcodes can also be used to characterize the ``topological regularity" of hydrogen-bonding networks. Figure \ref{fig:PE_Water} demonstrates the $\beta_1$ PEs for both TMAO and urea hydrogen-bonding networks at eight concentrations and two cutoff radii. Similar to molecular aggregation analysis, for each simulation, we consider 101 configurations or frames and generate 101 $\beta_1$ PEs. It can be seen that, the average PE value for both TMAO and urea hydrogen-bonding networks decreases with the concentration increase. The same pattern is observed at two local scales. Topologically, these results indicate that both hydrogen-bonding networks become more and more regular and lattice-like with concentration increase. Note that molecular aggregation has a totally different topological behavior, their PE value systematically increases with the concentration. More interestingly, the urea PE variance is significantly larger than that of TMAO and consistently increases with the concentration. This is exactly the same as in the urea aggregation systems.

In summary, we have used LPH models to explore the osmolyte molecular aggregation and their hydrogen-bonding networks. Essentially, we separate osmolyte molecules from water molecules, and study their local topological features separatively. In next section, we will focus on the interaction between osmolyte molecules and water molecules and characterize the topology of their interaction networks.

\subsection{IPH based topological features for osmolyte-water interaction network}\label{sec:IPH_results}

\begin{figure}
	\begin{center}
		\captionsetup[subfigure]{labelformat=empty,labelsep=none}
		\includegraphics[scale = 0.22]{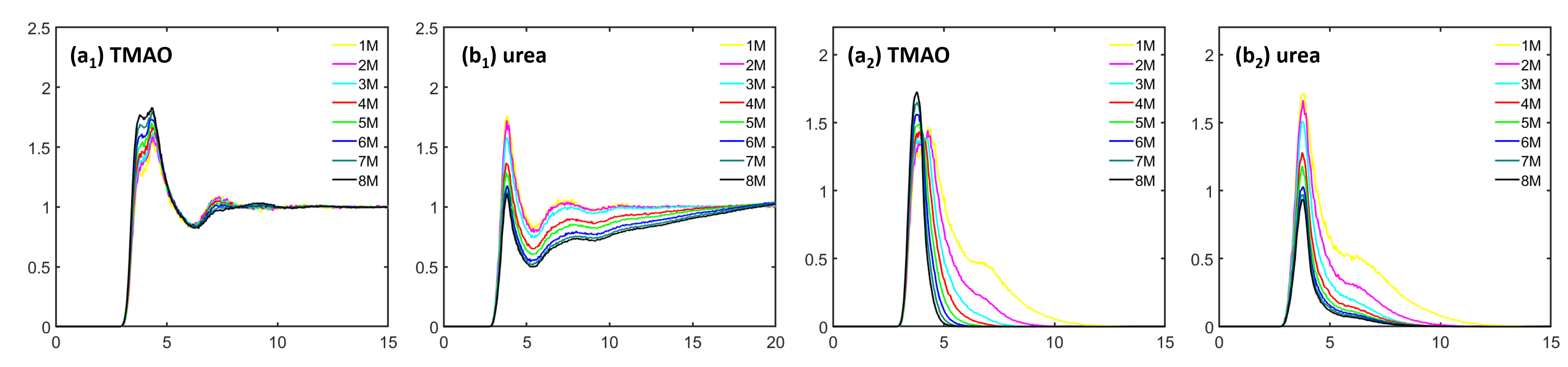}
		\caption{The comparison of global-scale and local-scale PRDFs for both TMAO and UREA. The PRDF of N-O is examined in the case of TMAO and C-O for the urea osmolyte. It can be seen that, the first peak value of the TMAO PRDFs increases with the concentration, while the first peak value of urea PRDFs decreases with the concentration.}
		\label{fig:DoubleLayer}
	\end{center}
\end{figure}

\begin{figure}
	\begin{center}
		\captionsetup[subfigure]{labelformat=empty,labelsep=none}
		\includegraphics[scale = 0.22]{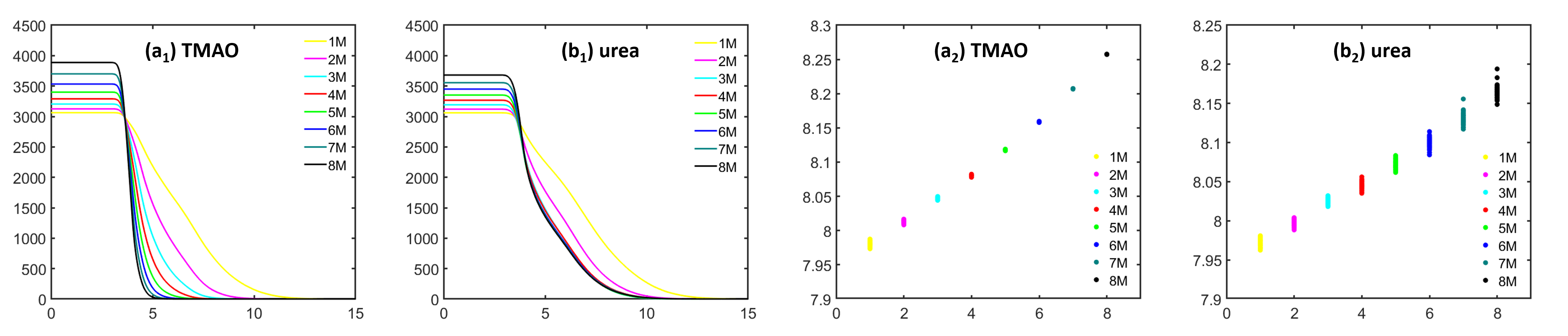}
		\caption{The comparison of average $\beta_0$ PBNs for hydrogen-bonding networks of TMAO and urea systems. The N-O pair interaction network is considered for analysis. The comparison of average $\beta_0$ PEs from the IPH analysis of TMAO and urea systems. For each configuration, a PE value can be calculated, thus a total 101 PEs are obtained for each simulation. It can be seen that, the average PE increases with the concentration for both TMAO and urea. However, the PE variance for urea systematically increases.}
		\label{fig:DoubleLayer_PBN}
	\end{center}
\end{figure}

In this section, we consider both global-scale and local-scale IPHs for analyzing interactions between osmolyte molecules and water molecules. In global-scale IPH, for each osmolyte molecule, we can construct an IPH matrix from Eq.(\ref{fig:IPH}) with the size $3001\times3001$, as there are totally 3000 water molecules. From the $\beta_0$ barcode of the IPH matrixes, a single PRDF can be calculated. And by averaging the PRDFs over all the 101 frames and all osmolyte molecules in each frame, we can obtain the average global-scale PRDF. In local-scale IPH, for each configuration or frame, an individual IPH can be constructed. Note that the size of the local-scale IPH matrix is $(N_{wtr}+N_{sml})\times(N_{wtr}+N_{sml})$, i.e., the total number of osmolyte and water molecules in the simulation. An average local-scale PRDF (or PBN) can be evaluated by averaging their values over all the 101 frames.

Figure \ref{fig:DoubleLayer} shows the comparison of global-scale and local-scale PRDFs for both TMAO and urea systems. {Both global-scale and local-scale PRDFs are normalized with the number density of the oxygen atom. The number density in global-scale is calculated by considering the number of oxygen atoms averaged over all spheres around each ion with radius $r_{max}$. The value of $r_{max}$ is half the box size. In the local-scale, the number density is simply the number of oxygen atoms divided by the volume of the simulation box for a given concentration.} Essentially, the global-scale PRDFs are identical to the traditional radial density functions. It can be seen that both TMAO and urea have two very obvious peaks, one located at around 4\AA~ and the other located at around 7\AA. However, their behaviors are dramatically different. For TMAO, the first peak value consistently increases with the concentration while the second peak value decreases with the concentration. The change of the TMAO peak values are relatively small, especially for the second peak. In contrast, both peaks of urea PRDFs vary greatly with concentration change. In local-scale IPH, PRDF values converge quickly to zero when the filtration value is larger than 12 \AA, which is dramatically different from the situation in global-scale PRDFs when their values converge to 1 at large filtration value. However, the first peak of local-scale PRDFs has similar pattern as the global-scale ones. The TMAO peak value increases with concentration, while urea peak value decreases with concentration. Moreover, at the region of filtration value from 5\AA~ and 10\AA, which is the region for the second peak of global PRDFs, the TMAO PRDF values decrease much faster than those of urea. When the concentration is larger than 5M, nearly all TMAO PRDF values drop to zero, while urea PRDF still remain largely positive.

To have a better understanding of the local-scale PRDFs, we check the PBNs and PEs from the local-scale IPH. Figure \ref{fig:DoubleLayer_PBN} demonstrates $\beta_0$ PBNs for TMAO and urea.  The $\beta_0$ PBNs are directly related to PRDFs. It can be seen that indeed the TMAO $\beta_0$ PBNs decrease much faster than those of TMAO at the filtration region from  5\AA~ to 10\AA, consistent with our observations in local-scale PRDFs. Further, we study the corresponding PEs. It can be seen in figure \ref{fig:DoubleLayer_PBN}, that the average PE values for both local-scale IPH models increase with the concentration. More interestingly, the PE variance for TMAO IPH decreases with the concentration, while that for urea systematically increases with the concentration.

%\begin{figure}
%	\begin{center}
%		\captionsetup[subfigure]{labelformat=empty,labelsep=none}
%		\includegraphics[scale = 0.4]{RDF_betti0_pe.png}
%		\caption{The comparison of average $\beta_0$ PEs from the IPH analysis of TMAO and urea systems. For each configuration, a PE value can be calculated, thus a total 101 PEs are obtained for each simulation. It can be seen that, the average PE increases with the concentration for both TMAO and urea. However, the PE variance for urea systematically increases.}
%		\label{fig:pe_newrdf}
%	\end{center}
%\end{figure}

\section{Conclusion}\label{sec:conclusion}

In this paper, we use the weighted persistent homology to study topological properties for osmolyte molecular aggregation and their hydrogen-bonding networks at a local scale. Two different models, i.e.,  localized persistent homology (LPH) and interactive persistent homology (IPH), are considered. We use persistent Betti number (PBN) and persistent entropy (PE) to quantitatively characterize the topological features from LPH and IPH. Based on persistent barcodes, we have proposed the persistent radial distribution function (PRDF). It has been found the the global-scale PRDF will reduce to traditional radial distribution function. While local-scale PRDFs can characterize the local interactions within the Voronoi cells. All these weighted persistent homology models can be used in any networks, graphs, biomolecules, etc.

\section*{Acknowledgments}
This work was supported in part by Nanyang Technological University Startup Grant M4081842 and Singapore Ministry of Education Academic Research fund Tier 1 RG31/18, Tier 2 MOE2018-T2-1-033.

\vspace{0.6cm}
\bibliographystyle{plain}
\bibliographystyle{unsrt}
%\bibliography{refs}

\end{document}